\documentclass[12pt]{iopart}
\usepackage{graphicx}
\usepackage{caption}
\usepackage{xcolor}

\usepackage{iopams} 
\expandafter\let\csname equation*\endcsname\relax
\expandafter\let\csname endequation*\endcsname\relax
\usepackage{amsmath}
\def\av#1{\langle #1 \rangle} 
\begin{document}

\title[]{Cluster size dependence of high-order harmonic generation}

\author{Y. Tao$^1$, R. Hagmeijer$^2$, H. M. J. Bastiaens$^1$, S. J. Goh$^1$, P. J. M. van der Slot$^1$, S. G. Biedron$^3$, S. V. Milton$^3$ and K. -J. Boller$^1$}

\address{$^1$ Laser Physics and Nonlinear Optics, Department of Science and Technology, MESA+ Institute for Nanotechnology, University of Twente, Enschede, The Netherlands

$^2$ Engineering Fluid Dynamics, University of Twente, Enschede, The Netherlands

$^3$ Department of Electrical and Computer Engineering, Colorado State University, Fort Collins, Colorado, USA}

\ead{steven95425@gmail.com}
\vspace{10pt}
\begin{indented}
\item[] 25$^{th}$ Nov 2016
\end{indented}

\begin{abstract}
We investigate high-order harmonic generation (HHG) from noble gas clusters in a supersonic gas jet. To identify the contribution of harmonic generation from clusters versus that from gas monomers, we measure the high-order harmonic output over a broad range of the total atomic number density in the jet (from $3\times10^{16}$ $\rm{cm}^{-3}$ to $3\times10^{18}$ $\rm{cm}^{-3}$) at two different reservoir temperatures (303 K and 363 K). For the first time in the evaluation of the harmonic yield in such measurements, the variation of the liquid mass fraction, $g$, versus pressure and temperature is taken into consideration, which we determine, reliably and consistently, to be below 20\% within our range of experimental parameters. By comparing the measured harmonic yield from a thin jet with the calculated corresponding yield from monomers alone, we find an increased emission of the harmonics when the average cluster size is less than 3000. Using $g$, under the assumption that the emission from monomers and clusters add up coherently, we calculate the ratio of the average single-atom response of an atom within a cluster to that of a monomer and find an enhancement of around 10 for very small average cluster size ($\sim200$). We do not find any dependence of the cut-off frequency on the composition of the cluster jet. This implies that HHG in clusters is based on electrons that return to their parent ions and not to neighbouring ions in the cluster. To fully employ the enhanced average single-atom response found for small average cluster sizes ($\sim200$), the nozzle producing the cluster jet must provide a large liquid mass fraction at these small cluster sizes for increasing the harmonic yield. Moreover, cluster jets may allow for quasi-phase matching, as the higher mass of clusters allows for a higher density contrast in spatially structuring the nonlinear medium.
\end{abstract}

%
\vspace{2pc}

\submitto{\NJP}
%
%
%
\section{Introduction}
Table-top sources based on high-order harmonic generation (HHG) provide coherent extreme ultraviolet (XUV) radiation on the femtosecond or even attosecond timescale~\cite{Salieres1995}. Such radiation is of great interest for various applications such as probing the ultrafast dynamics of atomic, molecular and solid systems~\cite{Uiberacker2007}, lensless diffractive imaging of objects at the nanoscale~\cite{Witte2014}, as well as seeding free-electron lasers~\cite{He2009}. Typically, noble gas atoms serve as the medium for HHG. On the single-atom level, the mechanism of such a process can be intuitively understood within a simple three-step model~\cite{Corkum1993,Lewenstein1994}: Initially, an electron escapes from its bound state in a strong drive laser field through tunnel ionization. Secondly, the electron is driven away and then accelerated back towards its parent ion. Finally, the electron recombines with its parent ion. However, in spite of progress with phase matching, the macroscopic output remains low~\cite{Constant1999, Hergott2002, Hadrich2014}. Recently, HHG from crystalline solid materials~\cite{Ghimire2011, Vampa2015a, Ndabashimiye2016} has been discovered and has shown a potential for higher conversion efficiency owing to the high density in solids. In addition, solids can be structured periodically on a micrometer scale which might further enable quasi-phase matching~\cite{Wu2015}. However, the generation mechanism of HHG in solids differs fundamentally from that in gas atoms. Unlike the atomic three-step model, knowledge of the complex electron dynamics inside the periodic structure of solids, responsible for the generation of harmonics, is missing. Moreover, in order to prevent permanent damage of the crystal, the drive laser intensities in those experiments are at least one order of magnitude lower than those conventionally applied in the gas medium, which results in a rather low cut-off energy~\cite{Ndabashimiye2016}.

Nanometer-sized clusters, formed via the van-der-Waals aggregation of gas atoms or molecules, provide an attractive alternative for HHG~\cite{Donnelly1996, Tisch1997, Vozzi2005, Ruf2013, Park2014, Aladi2014} since they combine the low average density of gas and the local high density of solids and liquids. This unique property should allow one to investigate the mechanism of HHG across the full range of relative densities from individual atoms up to solid materials. 
Furthermore, clusters have been shown as suitable to form spatially tailored density distributions that can be used for direct acceleration of particles~\cite{York2008}, an avenue that appears promising also for achieving quasi-phase matching in HHG.

However, the exact mechanism of HHG in clusters is not clear, e.g., it is not known to what extent the simple three-step model remains applicable for describing HHG in clusters. In particular, the recollision mechanism in the three-step model for clusters is not clarified yet. Several reports have attempted to improve the understanding of the mechanism of HHG based on their more detailed experimental observations. Donnelly et al.~\cite{Donnelly1996}, Vozzi et al.~\cite{Vozzi2005} and Aladi et al.~\cite{Aladi2016} observed an extended cut-off energy and enhanced conversion efficiency in clusters. The results suggest that the electron is initially tunnel ionized from one atom and later recollides with another neighbouring atom (atom-to-neighbour) instead of recombining with its parent atom (atom-to-itself). Such a mechanism would result in the generation of Bremsstrahlung (incoherent, broadband emission). Meanwhile, both Ruf et al.~\cite{Ruf2013} and Park et al.~\cite{Park2014} proposed another recollisional mechanism (cluster-to-itself). In that scenario, the harmonic radiation is assumed to be generated from a partially delocalized wave function spreading over the whole cluster. This is supported by measuring the ellipticity as well as the group delay of the high-order harmonics from clusters. An alternative recollision mechanism occuring in overdense plasmas~\cite{Quere2006} or solids~\cite{Ghimire2011} may also exist. Besides, resonant heating~\cite{Saalmann2006} mechanism can also occur during the tunnel ionization process.

In addition to these microscopic atomic-scale phenomena, HHG is a coherent emission process, such that the yield is also strongly affected by the macroscopic aspect, specifically phase matching, while further modifications can be caused by reabsorption of generated harmonics along the interaction length. Correspondingly, it is not easy to disentangle the single-particle (gas monomer or cluster) contribution from an experimental point of view. When attempting a measurement of the intrinsic (microscopic) nonlinear response of clusters versus their size, several considerations are of importance in an experiment. Generally, clusters are produced in a supersonic jet expansion of inert gas atoms. Both the average cluster size and density can be well controlled by the stagnation pressure and reservoir temperature~\cite{Dorchies2003, Smith1998}. However, when tuning these two experimental parameters, it is required to carefully keep the measured data, e.g., the generated harmonic order, out of ranges where strong phase mismatch and absorption limit or strongly influence the output signals. This is to ensure that the signals are large enough to be measureable by the detection system~\cite{Park2014}.

Except for the average cluster size and density, there is another important parameter, the liquid mass fraction, $g$, which characterizes the presence of clusters. This parameter is defined as the ratio of the number of atoms in the form of clusters to the total number of atoms in the jet. For most of the experiments mentioned above~\cite{Tisch1997, Vozzi2005, Park2014, Aladi2014}, the researchers interpret their results by choosing $g$=1 without further justification, namely, they assume that a pure cluster jet is generated and thereby the measured high-order harmonic (HH) signals are \emph{entirely} to be attributed to clusters. However, both our recent modelling of cluster formation~\cite{Tao2016} and other measurements~\cite{Gao2012, Gao2013} strongly indicate that $g$ is not unity but dependent on both the stagnation pressure and reservoir temperature. For instance, for our slit nozzle, the value of $g$ for argon clusters lies only at about $20\%$ at room temperature over a broad range of stagnation pressures. Even at very low reservoir temperatures ($\sim 173$ K, via cooling by pre-cooled nitrogen gas~\cite{Gao2013}), $g$ only rises up to $\sim40\%$. As a result, the assumption of $g=1$ misleads the interpretation of the measurements. For a valid determination of the nonlinearity of clusters, one has to take into account the contribution to the HH yield from both clusters as well as gas monomers when interpreting experimental data.

In this work, we present a detailed experimental study on HHG from a supersonic argon jet within a similar range of the total atomic number density as in previous studies (from $\sim10^{16}$ to $10^{18}$ $\rm{cm^{-3}}$)~\cite{Vozzi2005, Ruf2013, Park2014, Aladi2014}.  However, in order to better clarify a possible dependence of HHG on the average cluster size, we change the stagnation pressure over a broad range to maximize the variation in cluster size. Importantly, for disentangling the contribution to HHG from clusters and gas monomers, we perform experiments at two different reservoir temperatures in order to vary the liquid mass fraction, $g$, for the same range of cluster sizes. We determine the dependence of the liquid mass fraction, $g$, on both stagnation pressure and reservoir temperature with a high degree of reliability using our one-dimensional model~\cite{Tao2016}. We find that about a maximum of 20\% of the gas atoms are converted into clusters under our experimental conditions. Comparing the measured HH yield at the $21^{st}$ harmonic order for the cluster jet with the corresponding calculated yield from monomers only, we find an enhanced harmonic yield when the average cluster size, $\av{N}$, is within $200\lesssim\av{N}\lesssim3000$. We use the liquid mass fraction, $g$, to calculate the relative average single-atom response for an atom inside a cluster to that of a monomer. We find an enhancement of up to a factor of 10 when $\av{N}\lesssim 500$. This result is in agreement with earlier work~\cite{Park2014} that concludes an increased nonlinearity of atoms in clusters for sufficiently small average cluster size, however, here we quantify the enhancement factor vs. the average cluster size.
At the same time, we observe no changes in the cut-off energy when the average cluster size increases. This confirms other reports~\cite{Park2014} suggesting that the collision mechanism described in the three-step model for HHG in gas atoms may still be applicable for HHG in clusters. This means that the tunneled electron recombines only with its parent ion.

\section{Experimental setup}
\begin{figure}[ht!]
\centering
\captionsetup{width=0.8\linewidth}
\includegraphics[width=0.8\textwidth]{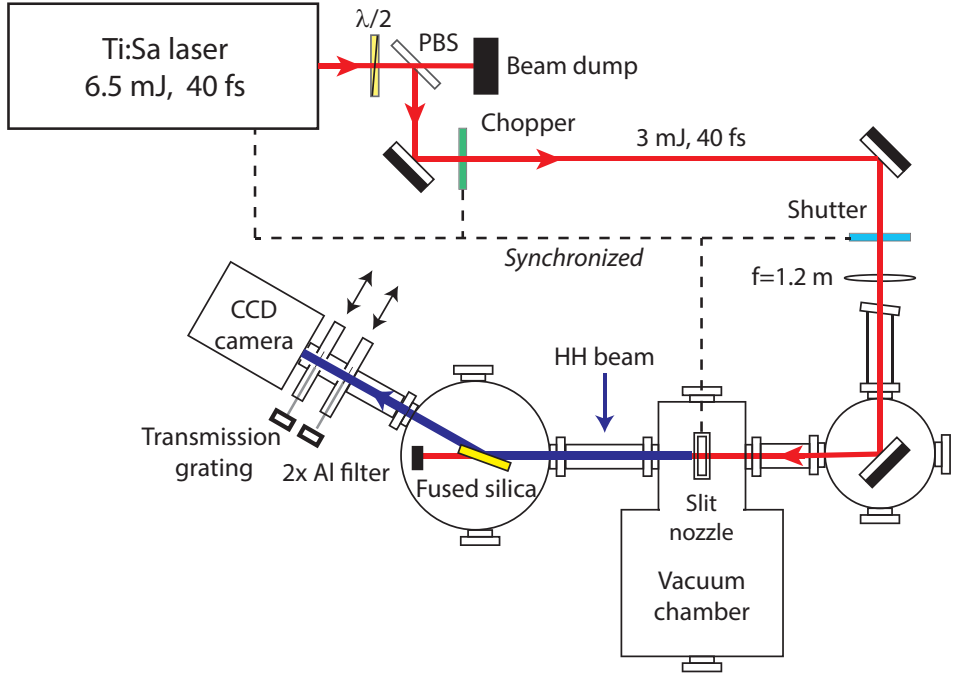}
\caption{Schematic of the experimental setup for HHG in an argon supersonic jet.}
\label{fig: Experimental_setup}
\end{figure}

The experimental setup used for HHG in clusters is depicted in Fig. \ref{fig: Experimental_setup}. Argon clusters are generated from a supersonic slit nozzle (rectangular cross section of exit: 1.0$\times$5.0 $\rm{mm^2}$, expansion half-angle: $14\,^{\circ}$) mounted on top of an electro-magnet driven pulsed gas valve (Parker, 9 series) inside a vacuum chamber. The stagnation pressure applied to the valve can be varied between 0 and 70 bar, with an accuracy of 0.2 bar at low pressures (0 to 5 bar), and with an accuracy of 0.5 bar in the higher pressure regime (5 to 70 bar). With a resistive heater, the temperature of the nozzle can be adjusted between room temperature and $105\,^{\circ}\rm{C}$ with an accuracy of about $0.5\,^{\circ}\rm{C}$. The average cluster size and density, as well as the total atomic number density generated with the supersonic nozzle were measured as a function of stagnation pressure at different temperatures as presented in our previous published paper~\cite{Tao2016}. To exclude the influence of changing phase matching conditions in the build-up of the harmonic field, the slit nozzle is oriented perpendicular to the drive laser beam, such that the laser beam propagates through the short dimension (width) of the jet ensuring that the interaction length is shorter than the coherence length.  
For driving HHG, we employ a femtosecond Ti:Sapphire laser system operating at a center wavelength of 795 nm at 1 kHz repetition rate (Legend Elite Duo HP USP, Coherent Inc.). The laser generates linearly polarized output pulses with a maximum pulse energy of 6.5 mJ and a pulse duration of about 40 fs~\cite{Goh2016}. 
To avoid any major self-phase modulation and plasma defocusing along the propagation direction, the pulse energy used in our experiment is limited to a fixed value of 3.0 mJ via a variable attenuator comprising a rotatable half-wave plate followed by a thin film polarizing beam splitter (PBS).
The laser pulse is loosely focused about 1 mm above the nozzle using a lens of 1200 mm focal length, resulting in a peak intensity of about $1.5\times10^{14}$  $\rm{W/cm^{2}}$ at focus and an effective interaction length~\cite{Tao2016} of about $650$ $\rm{\mu}$m. The relatively low peak intensity and short pulse duration ensures that the harmonic emission involves only clusters that are not already affected by ionization induced explosion and disintegration, since the time scale for these processes ($\sim$ hundreds of femtoseconds) is much longer than the drive laser pulse duration ($\sim$ 40 fs)~\cite{Park2014}.
The generated HH beam co-propagates with the drive laser beam, from which it is primarily separated by an uncoated fused silica plate placed at an incident angle of 75 degree. This incident angle is close to the Brewster angle for the drive laser beam (at center wavelength), such that most of the drive laser beam is transmitted and diverted to an absorbing beam dump. Any residual drive laser radiation that is reflected from the fused silica plate is fully blocked by a set of two 200-nm thick aluminium (Al) filters placed in series. These filters act as a band pass filter for the harmonic radiation, transmitting more than 40\% in the wavelength range of 17 to 80 nm~\cite{Goh2015}. The transmitted HH beam is detected by an XUV CCD camera (Princeton Instruments, PIXIS-XO 2048B) placed behind the filters. For the measurement of the spectral distribution of the HH output, an in-house fabricated  transmission grating (3,000 lines/mm)~\cite{Goh2015} illuminated through a 300 $\rm{\mu{m}}$ slit is shifted into the beam path. To minimize absorption of HH radiation in the beam path towards the detection system, the pulsed gas valve is operated at a low repetition rate of 1 Hz, as to keep the background pressure below $10^{-3}$ mbar during operation. A mechanical chopper system (MC20008B-EC, Thorlabs Inc) is inserted into the beam path, which reduces the repetition rate of the drive laser from 1 kHz to 71 Hz, in order to prevent any damage to the fused silica plate and the Al filters due to high average power. For allowing the single-shot detection of the HH beam profile, an additional mechanical shutter (SH05, Thorlabs Inc) could be inserted into the beam path, reducing the repetition rate of the laser further to 1 Hz.
\section{Result and Discussion}
As was emphasized above, it is essential for correct data interpretation that the influence of the liquid mass fraction, $g$, on the HH yield is clarified, because only this enables us to resolve the relative contribution from clusters and gas monomers. 
Here, we first determine the dependence of the liquid mass fraction, $g$, on the two main experimental parameters, which are the stagnation pressure, $p_0$, and the reservoir temperature, $T_0$. An understanding of this dependence is required for the analysis of the measured HH intensity with the stagnation pressure (which determines the atomic number density) as shown below.
Next, we discuss the complete HH spectra and the cut-off wavelength measured for two specific temperatures, at three different stagnation pressures. We then focus on the strongest harmonic (HH21) and discuss the yield as a function of total atomic number density for two different temperatures, as well as the calculated variation in absorption and coherence length. We use a simple theoretical model to estimate the yield for HH21 from the monomers, which includes absorption and phase matching effects. Using this model in combination with the liquid mass fraction, we analyze the relative contribution of clusters and gas atoms to the HH yield. From this, we obtain the single-atom response for atoms inside clusters with different sizes. 

\begin{figure}[ht!]
\centering
\captionsetup{width=0.8\linewidth}
\includegraphics[width=0.8\textwidth]{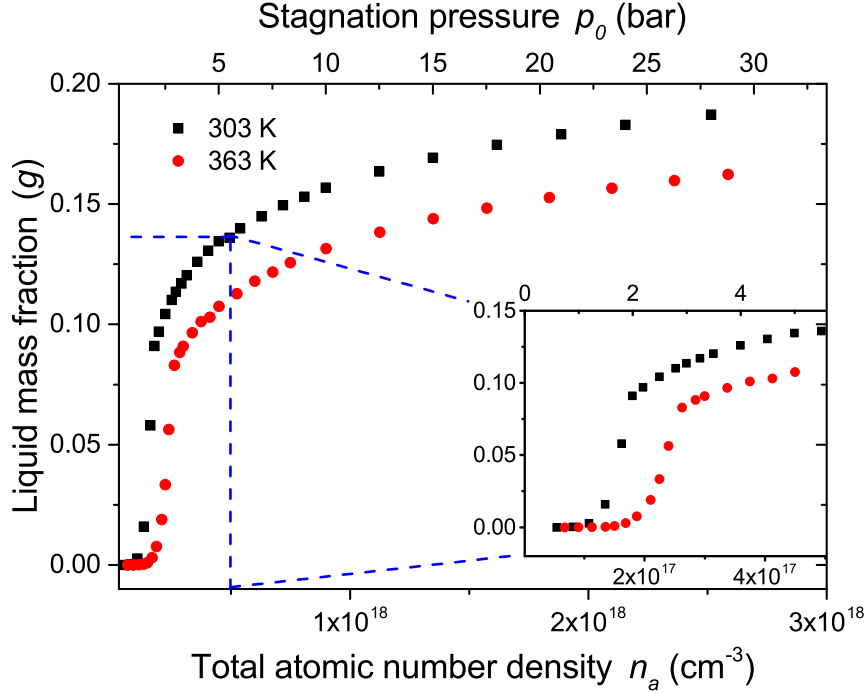}
\caption{Liquid mass fraction, $g$, in the supersonic argon jet as a function of the total atomic number density, $n_a$, for two reservoir temperatures $T=303$~K (black squares) and $T=363$~K (red circles), as obtained from Ref~\cite{Tao2016}.}
\label{fig: liquid_mass_fraction}
\end{figure}

Figure~\ref{fig: liquid_mass_fraction} shows the liquid mass fraction, $g$, across a broad range of stagnation pressures, $p_0$, (from 300 mbar to 35 bar), and the two reservoir temperatures $T=303$~K (black squares) and $T=363$~K (red circles) used in the HHG experiments. To determine $g$, we have used the relation that we derived previously by combining interferometry and Rayleigh scattering data with a theoretical description of cluster formation~\cite{Tao2016}. 
To allow a direct comparison of the liquid mass fraction at different temperatures, the stagnation pressure is converted into the total atomic number density, $n_a\propto{p_0/T_0}$, calibrated by an interferometric measurement~\cite{Tao2016}.
It can be seen that the liquid mass fraction is far off unity in that it grows from an extremely small value to a maximum of about 19\% as $n_a$ is varied from $\sim0.5\times10^{17}$ to $\sim 2.5\times10^{18}$~cm$^{-3}$.
The inset shows an enlarged view of the growth of $g$ in the low-density region up to $n_a=5\times10^{17}\;\rm{cm^{-3}}$.
From Fig. \ref{fig: liquid_mass_fraction}, it can be clearly seen that, at $T_1=303$ K (black squares), the liquid mass fraction is very small ($\leq 0.01$) for densities up to $n_a=10^{17}\;\rm{cm^{-3}}$, and increases rapidly up to 10\% at a density of about $n_a=1.8\times10^{17}\;\rm{cm^{-3}}$. Above this density, $g$ grows more weakly, reaching its maximum value of about 19\% for a density near $n_a=2.5\times10^{18}\;\rm{cm^{-3}}$. The growth trend of the liquid mass faction at increased temperature, $T_2=363$ K (red circles), is very similar, although setting in at a higher density of about $n_a=2\times10^{17}\;\rm{cm^{-3}}$. In this case, $g$ reaches a value of 16\% for a density around $n_a=2.5\times10^{18}\;\rm{cm^{-3}}$.
Within the entire range of the densities and temperatures accessible in our experiment, we find that the liquid mass fraction remains lower than 19\%.
We note that these rather low values for $g$ correspond well with the experimental results reported by others~\cite{Gao2013, Shim2007}. Our modelling results show that even at cryogenic temperatures (at 170 K, which can be achieved using pre-cooled nitrogen gas) and high stagnation pressures (50 bar), the maximum liquid mass fraction still does not rise above 40\%. From this we conclude that all the previous experimental investigations aiming to unravel the size dependent cluster contribution to the HH yield from the supersonic gas jet have provided questionable conclusions as it is not justified to neglect the major presence of atoms as monomers~\cite{Vozzi2005, Park2014}.

To study more quantitatively the influence of the liquid mass fraction on the HH yield, we raise the temperature of the nozzle to decrease $g$, while keeping the total atomic number density in the jet unchanged. This is accomplished by increasing the stagnation pressure such that the ratio $p_0/T_0$ remains constant.
\begin{figure}[ht!]
\centering
\captionsetup{width=0.8\linewidth}
\includegraphics[width=0.8\textwidth]{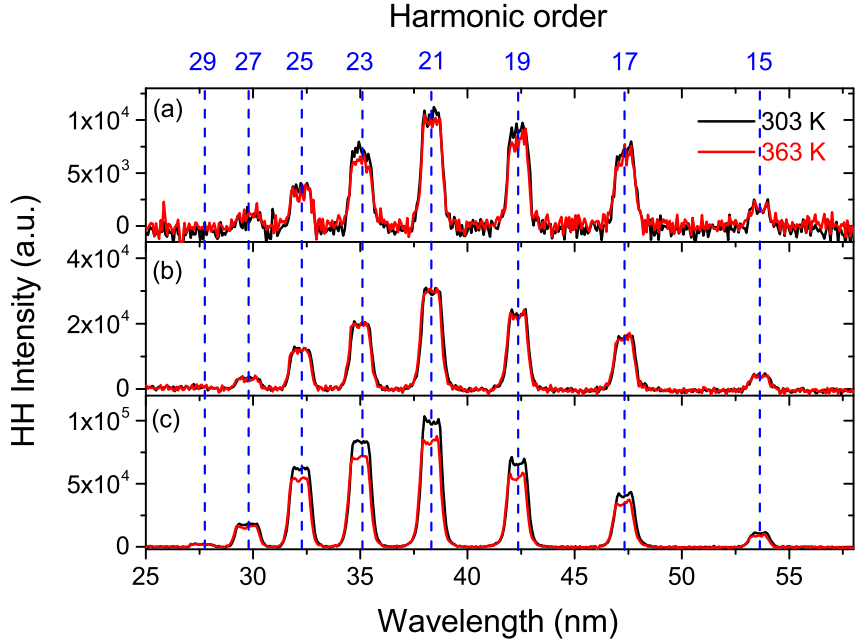}
\caption{HH spectra measured at 303~K (black traces) and 363~K (red traces) for the total atomic number densities, $n_a\approx$ $1.6\times{10^{17}}$ (a), $4.5\times{10^{17}}$ (b) and $2.1\times{10^{18}}$ $\rm{cm^{-3}}$ (c). The quasi-flat-top shape of the peaks is caused by the limited resolution of the spectrometer, chosen for maximizing the output signal.}
\label{fig: spectrum_comparison}
\end{figure}
In Fig. \ref{fig: spectrum_comparison}, we present a series of six typical HH spectra measured at two specific temperatures (303~K (black traces) and 363~K (red traces)), for three different total atomic number densities, $n_a$, falling within the low (a), middle (b) and high-density regions (c) of the experimental measurement range. Each spectrum is integrated over 100 laser shots to increase the signal-to-noise ratio as well as reduce the influence due to the fluctuation of the drive laser pulse energy (typically $\sim$5\%). We note that the quasi-flat-top shape of the peaks is due to the relative large slit width set to maximize the harmonic output signal.
For measuring the HH spectra in the low-density region, the density is set to around $1.6\times{10^{17}}$ $\rm{cm^{-3}}$ (with $p_0$=1.5~bar and $T=303$ ~K, or with $p_0$=1.8~bar and $T=363$~K). In the middle and high-density regions, the densities are set to around $4.5\times{10^{17}}$ $\rm{cm^{-3}}$ (with $p_0$=5.0 bar and $T=303$~K, or with $p_0$=6.0 bar and $T=363$~K) and to around $2.1\times{10^{18}}$ $\rm{cm^{-3}}$ (with $p_0$=24 bar and $T=303$~K, or with $p_0$=28 bar and $T=363$~K).
From Fig. \ref{fig: spectrum_comparison}, it can be seen that the spectra comprise in total eight harmonic orders, ranging from the $15^{\rm{th}}$ to the $29^{\rm{th}}$. Note that the 29$^{\rm{th}}$ harmonic intensity in the low-density region is so weak that it is not observable in all spectra. Among these harmonic orders, the $21^{\rm{st}}$ harmonic consistently exhibits the strongest intensity. Another feature observable in Fig. \ref{fig: spectrum_comparison} is that the intensity of the harmonics grows with the increasing total atomic number density, from the low-density region to the high-density region. The HH intensity measured at $T=303$ K is found to be very similar to that measured at $T=363$ K in both the low and the middle-density regions, while it is slightly higher in the high-density region. Remarkably, we find that the relative shape of the spectra remains identical, independent of the total atomic number density and the temperature. This suggests that the two-field combinations originating from clusters on one hand and gas monomers on the other hand are emitted coherently, i.e., without any change in relative phase, when the density is increased.
On the long wavelength side, the spectra are limited to about 53~nm (15$^{\rm{th}}$ order). This limit can be traced back to strong reabsorption of the generated XUV radiation in argon~\cite{Durfee1999}. On the short wavelength side, the spectra are limited to about 28~nm (29$^{\rm{th}}$ order). This wavelength agrees well with the calculated cut-off wavelength according to the three-step model ($\lambda_{\rm{cutoff}}=hc/(I_p+3.17U_p)\approx28\;\rm{nm}$, where $U_p$ is calculated from the experimental laser parameters). Nevertheless, the measured cut-off wavelengths could imply different mechanisms acting in the recombination process during HHG. Specifically, for HHG in cluster jets, a huge extension of the cut-off wavelength towards shorter wavelengths was previously observed, which has been explained by the so-called atom-to-neighbour collision mechanism~\cite{Vozzi2005, Aladi2014, Zaretsky2010}. Our observation of the cut-off wavelength does not show such an extension in the measured wavelength range. Instead, the cut-off coincides with the predicted value from the three-step model, as confirmed by experiments of others, suggesting that the coherent emission from clusters is due to the recombination of the tunnel-ionized electron with its parent ion within the cluster (atom-to-itself collision mechanism)~\cite{Ruf2013, Park2014}. 

Further investigating the high-order harmonic contribution from clusters and gas monomers, we record the harmonic spectra over a broad range of the total atomic number density, $n_a$, from $n_a=6\times10^{16}\;\rm{cm^{-3}}$ to $n_a=2.5\times10^{18}\;\rm{cm^{-3}}$. Since the relative HH intensity distribution in the measured spectra does not change with density (see Fig. \ref{fig: spectrum_comparison}), it allows us to select, as an example, the $21^{\rm{st}}$ harmonic ($\approx38$ nm) in the spectra as representative also for the other harmonic orders. The motivation of choosing the $21^{\rm{st}}$ harmonic is that it provides the strongest signal in all the spectra and therefore provides the best signal-to-noise ratio. 
\begin{figure}[ht!]
\centering
\captionsetup{width=0.8\linewidth}
\includegraphics[width=0.8\textwidth]{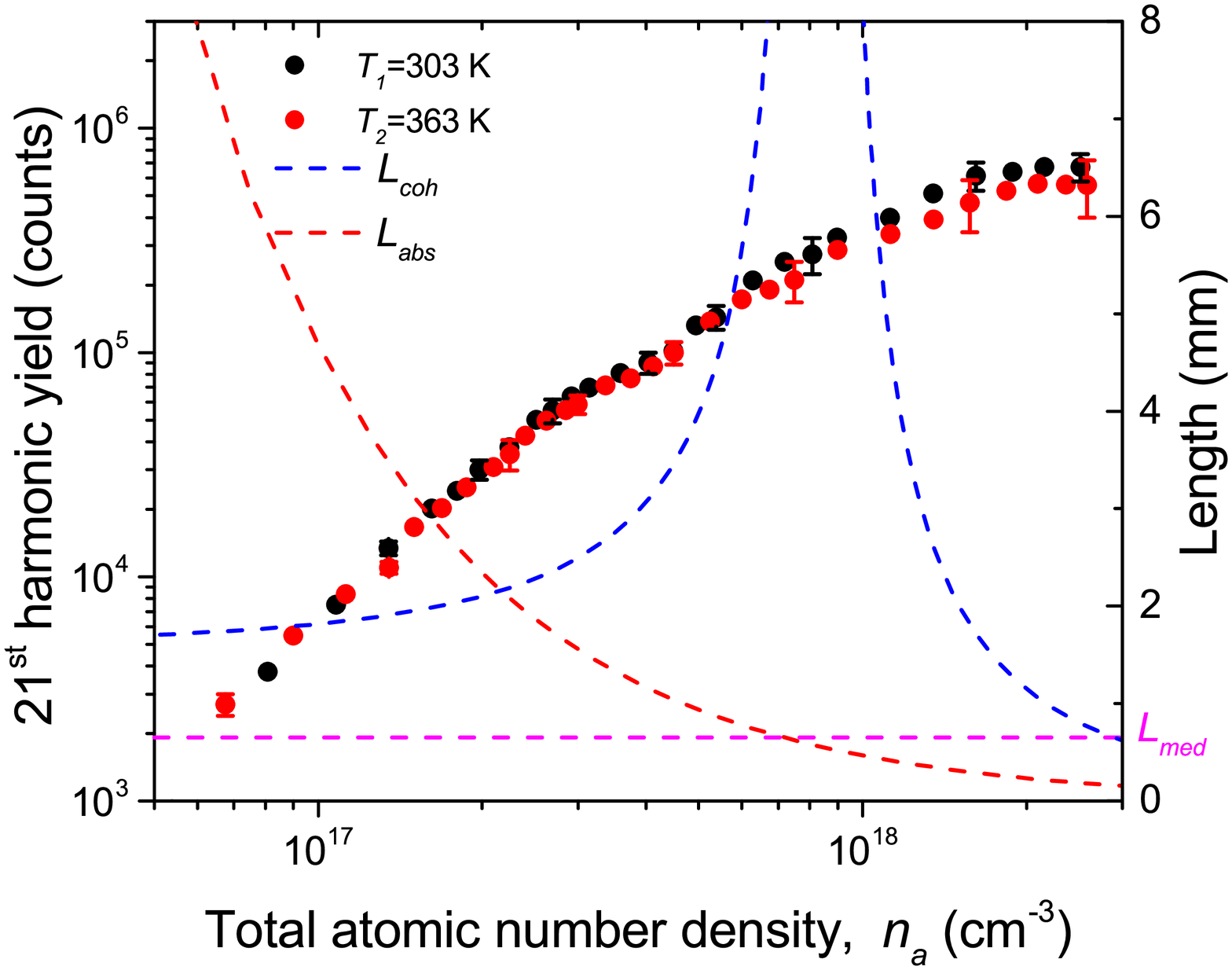}
\caption{The average yield (over 100 shots) at the $21^{\rm{st}}$ harmonic as a function of the total atomic number density, $n_a$, for two different reservoir temperatures, 303~K (black) and 363~K (red) respectively. Also shown are the calculated coherence length (blue dashed curve), absorption length (red dashed curve) for the $21^{\rm{st}}$ harmonic as well as the medium length ($L_{med}=0.65$ mm, pink dashed line).}
\label{fig: HH21intensity_vs_na}
\end{figure}
In Fig.~\ref{fig: HH21intensity_vs_na}, we plot the average $21^{\rm{st}}$ harmonic yield obtained by spectrally integrating the $21^{\rm{st}}$ harmonic signal in the spectra, versus the total atomic number density, $n_a$, for two different temperatures on a double logarithmic scale. For each measurement, the HH yield at the $21^{\rm{st}}$ harmonic is integrated over 100 shots. As the single-shot HH yield at a specific order is weak, especially for the measurements at low densities, the error bars shown in Fig.~\ref{fig: HH21intensity_vs_na} are derived from the shot-to-shot fluctuations in the measurement of the total harmonic beam energy for a total of 100 different shots, assuming that the relative error in the average spectral intensity is equal to the relative error in the total harmonic beam energy.
From Fig.~\ref{fig: HH21intensity_vs_na}, it can be seen that the $21^{\rm{st}}$ harmonic yield is almost the same for both temperatures within the experimental fluctuation of the harmonic yield ($\sim$10\%). he yield initially grows as $n_a^2$ at small $n_a$ ($\lesssim 1.5\times10^{17}$~cm$^-3$) and slows down for larger $n_a$. We also notice that the HH yield summed up over all harmonic orders in the spectra, i.e., from the $15^{\rm{th}}$ harmonic to the $27^{\rm{th}}$ harmonic, gives a similar growth trend.

To determine the effect of absorption and phase matching on the HH yield when $n_a$ is increased, we plot the calculated absorption length~\cite{Henke1992} ($L_{abs}$, red dashed curve) and the coherence length ($L_{coh}$, blue dashed curve) in Fig. \ref{fig: HH21intensity_vs_na} versus the atomic number density together with the effective experimental interaction length (effective length of the medium, $L_{med}=0.65$ mm, pink dashed line as determined in Ref~\cite{Tao2016}).
The absorption length, starting with a rather big value, $L_{abs}=6.5$ mm, drops gradually with increasing $n_a$. The length becomes smaller than the length of the medium at higher densities ($n_a\geq7\times10^{17}\;\rm{cm^{-3}}$), which means that here the measured $21^{\rm{st}}$ harmonic yield is mainly limited by reabsorption in the jet. That absorption does not play a role at the lower densities (e.g., $n_a\lesssim4\times10^{17}\;\rm{cm^{-3}}$) is experimentally verified by comparing the growth of the HH yield for different orders, in particular for the $15^{\rm{th}}$ and the $27^{\rm{th}}$ harmonic, as the absorption length is strongly wavelength dependent. No difference in growth was observed.

The coherence length, $L_{coh}$, is calculated from the wave-vector mismatch resulting from the atomic, the plasma, and the geometric dispersion~\cite{Kazamias2011}. It can be noticed that $L_{coh}$ remains at least three to four times longer than $L_{med}$ for a broad range of densities. In the range from $n_a=5\times10^{16}\;\rm{cm^{-3}}$ to $n_a=3\times10^{17}\;\rm{cm^{-3}}$, the wave-vector mismatch is mainly dominated by the geometry dispersion originating from the Gouy phase shift, which is independent of $n_a$.
Further increasing $n_a$, $L_{coh}$ grows rapidly and reaches its maximum value around $n_a=8\times10^{17}\;\rm{cm^{-3}}$ where the minimum wave-vector mismatch is achieved. For $n_a>8\times10^{17}\;\rm{cm^{-3}}$, $L_{coh}$ drops dramatically due to the dispersion of a large density of ionized electrons and becomes even shorter than $L_{med}$ around $n_a=2.5\times10^{18}\;\rm{cm^{-3}}$. We note that as $L_{coh}$ is at least a factor of 3 larger than $L_{med}$ over the range of interest for $n_a$, it does not have a strong effect on the HH yield and may be responsible for the small oscillation visible in the HH yield around $n_a\approx7\times10^{17}$~cm$^-3$, as can be seen in Fig.~\ref{fig: HH21intensity_vs_na}.

In order to separate the contribution from the clusters to the HH yield from that of the argon monomers, we need to take into consideration both the liquid mass fraction, $g$, and the monomer contribution. For the latter we use the basic approach of Durfee et al.~\cite{Durfee1999} and Constant et al.~\cite{Constant1999}, which includes both absorption and phase matching. Within this model, the number of photons, $N_{21}$, at the $21^{\rm{st}}$ harmonic frequency emitted on axis is proportional to 
\begin{equation}
n_a^2 A_{m,21}^2 C(L_{abs},L_{coh},L_{med}), 
\label{eq:Nout}
\end{equation}	
where $A_{m,21}$ is the atomic response of monomers at the $21^{\rm{st}}$ harmonic and $C(L_{abs},L_{coh},L_{med})$ is a system dependent constant that depends on the absorption, coherence and effective medium length and is therefore also dependent on the total atomic number density, $n_a$.
\begin{figure}[ht!]
	\centering
	\captionsetup{width=0.8\linewidth}
	\includegraphics[width=0.74\textwidth]{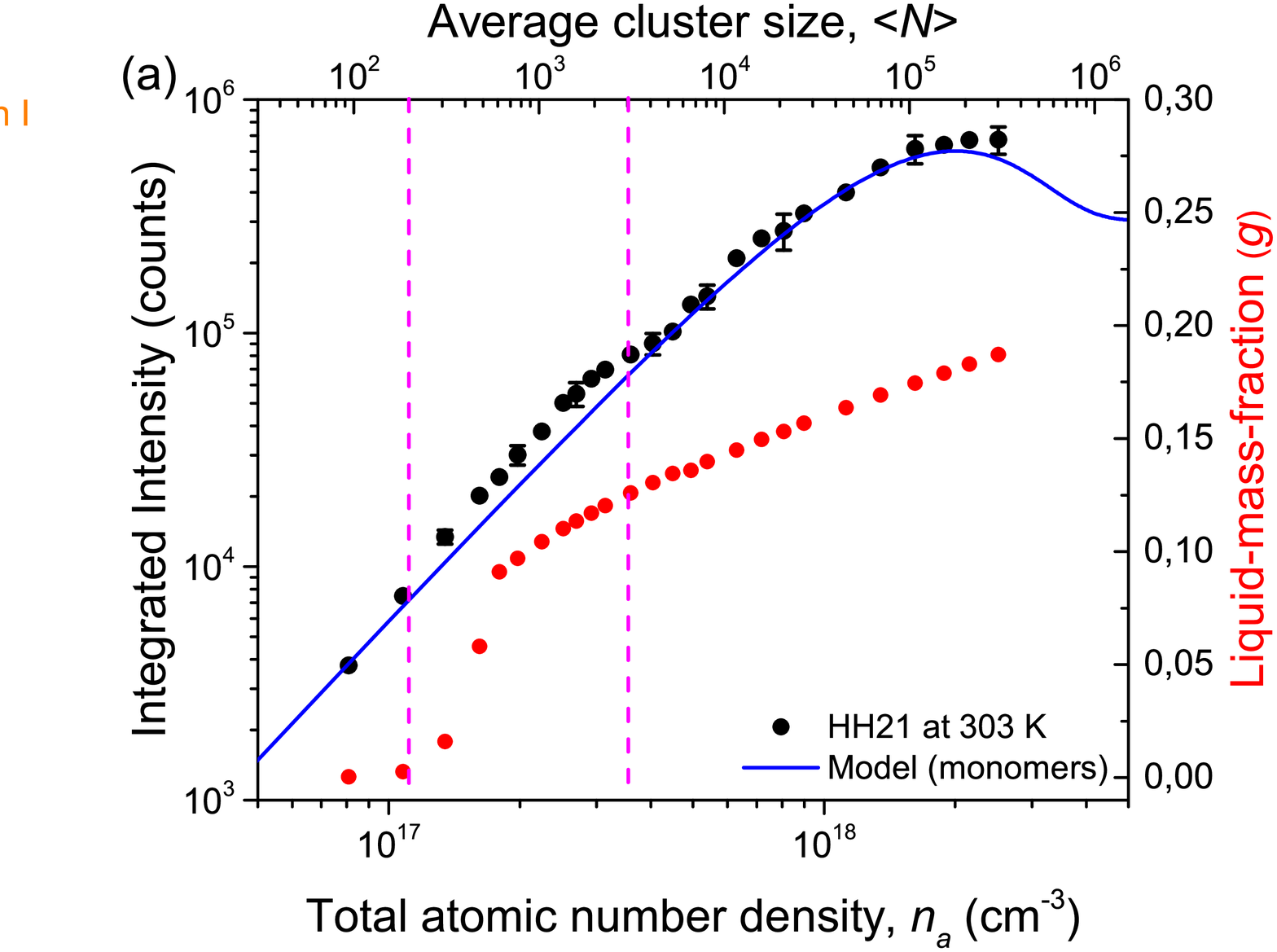}
	\includegraphics[width=0.74\textwidth]{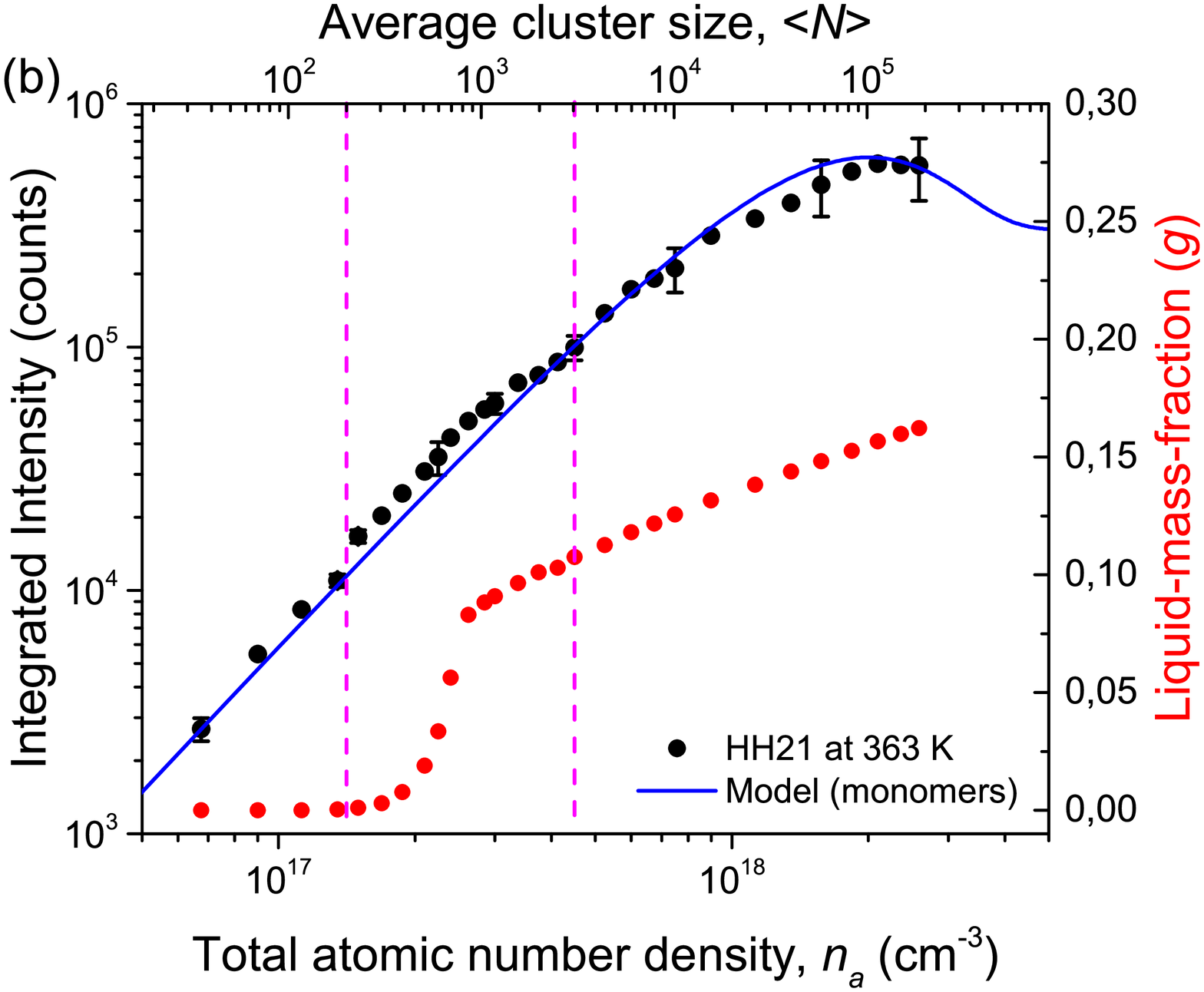}
	\caption{The measured $21^{\rm{st}}$ harmonic yield for the cluster jet  (black circles), calculated $21^{\rm{st}}$ harmonic yield  for pure monomers (blue line) and the liquid mass fraction, $g$, (red circles) versus the total atomic number density, $n_a$, at two different temperatures ($T=303$ K (a) and $T=363$ K (b)). The corresponding average cluster size, $\av{N}$, is displayed at the top axes. The pink dashed lines indicate the range for which the relative single-atom response for atoms in clusters has been calculated.}
	\label{fig: HH21intensity_vs_na_temp}
\end{figure}
In Fig.~\ref{fig: HH21intensity_vs_na_temp} we replot the measured HH yield for the $21^{\rm{st}}$ harmonic of the cluster jet (black circles) together with the liquid mass fraction, $g$, (red circles) and the calculated HH yield for pure monomers (solid blue line) as a function of the total atomic number density, $n_a$, for a temperature of 303~K (Fig.~\ref{fig: HH21intensity_vs_na_temp}(a) and 363~K (Fig.~\ref{fig: HH21intensity_vs_na_temp}(b)). 
As the measurements only provide a relative yield, the calculated yield produced by pure monomers for the $21^{\rm{st}}$ harmonic is scaled to the measured value at the lowest $n_a$, where the nozzle only produces monomers, i.e., the liquid mass fraction is very small. We note that only a single scale factor was required to match the model calculation to the measured value at the two temperatures.  At each of the two temperatures, the average cluster size, $\av{N}$, that corresponds to the total atomic number density, $n_a$, is displayed on the top axis. $\av{N}$ is determined from the stagnation pressure, $p_0$, and the temperature, $T$, using Hagena's law (for $\av{N}<1000$)~\cite{Hagena1992}  and the newly developed power law (for $\av{N}\geqslant1000$) from Tao et al.~\cite{Tao2016}. Figure~\ref{fig: HH21intensity_vs_na_temp} shows that the measured yield at the $21^{\rm{st}}$ harmonic agrees within measurement accuracy with the corresponding calculated yield for pure argon monomers when  $\av{N} \gtrsim3000$. Therefore, we do not observe any enhanced emission of the $21^{\rm{st}}$ harmonic in this regime. Nonetheless, when $\av{N}\lesssim 3000$, Fig.~\ref{fig: HH21intensity_vs_na_temp} shows for both temperatures a slightly higher yield for the $21^{\rm{st}}$ harmonic for the cluster jet compared to corresponding yield from monomers only. This indicates that atoms in small clusters with average size $\av{N}\lesssim3000$ are more efficient in emitting HH than monomers. This may even be true down to a very small average cluster size, however, the lower limit is masked in our experiment by the small liquid mass fraction that is produced by our nozzle at the lowest total atomic number density.

In the range $200\lesssim\av{N}\lesssim3000$, as indicated by the dashed pink lines, Fig.~\ref{fig: HH21intensity_vs_na_temp} shows a slightly larger $21^{\rm{st}}$ harmonic yield for the cluster jet compared to that from pure monomers with the same $n_a$. For a given $n_a$, the value of $g$ determines the fraction of atoms in the form of clusters with a size distribution having an average size, $\av{N}$, that may generate harmonics with a different efficiency than the efficiency of the gas monomers. To derive the single-atom response for the clusters relative to that of gas monomers, we propose a simple model that relies on the liquid mass fraction, $g$, to separate the contribution from clusters and monomers to the total yield at a given total atomic number density, $n_a$. Note that, the liquid mass fraction, $g$, was misrepresented in previous experiments by simply assuming $g=1$. 
First, when the medium consists of only monomers, the yield $N_{m,21}$ is given by Eq.~\eqref{eq:Nout}. Second, since $L_{abs}$ and $L_{coh}$ depend on the total atomic number density, $n_a$ and $L_{med}<<L_{coh}$, we can assume that the macroscopic medium response represented by the factor $C$ in eq.~\eqref{eq:Nout} is the same for the medium consisting of pure monomers or a monomer and cluster mixture. Finally, for the cluster jet with $g>0$, the medium consists of a mixture of clusters with average size, $\av{N}$, and monomers and Eq.~\eqref{eq:Nout} needs to be extended. Let $A_{c,21}(\av{N})$ be the single-atom response of atoms within clusters with average size $\av{N}$. Assuming that the harmonic field produced by the monomers and cluster add up coherently as described above, the yield, $N_{c,21}$, for the cluster and monomer mixture is proportional to
\begin{equation}
n_a^2 [(1-g(\av{N},T))A_{m,21} + g(\av{N},T)A_{c,21}(\av{N})]^2 C(L_{abs},L_{coh},L_{med}),
\label{eq:Nout-jet}
\end{equation} 
where it is made explicit that the liquid mass fraction, $g$, is a function of the temperature, $T$, and the average cluster size, $\av{N}$. As the calculated yield produced by monomers only is scaled to the experimentally determined yield produced by the monomer and cluster mixture, the proportional factor is the same, which takes into account the efficiency with which the harmonic radiation emitted from the jet reaches the XUV CCD camera, including the reflectance of the fused silica plate, the transmission of the Al filters and the grating, as well as the responsivity of the camera. Hence, the ratio $N_{c,21}$ to $N_{m,21}$ is given by
\begin{equation}
\frac{N_{c,21}}{N_{m,21}}=\frac{(1-g(\av{N},T))A_{m,21} + g(\av{N},T)A_{c,21}(\av{N})]^2}{A_{m,21}^2},
\end{equation}  
which can be rewritten into
\begin{equation}
\frac{A_{c,21}(\av{N})}{A_{m,21}}=\frac{1}{g(\av{N},T)}\left(\sqrt{\frac{N_{c,21}}{N_{m,21}}}+g(\av{N},T)-1\right).
\label{eq:ratio}
\end{equation}  

\begin{figure}[ht!]
\centering
\captionsetup{width=0.8\linewidth}
\includegraphics[width=0.75\textwidth]{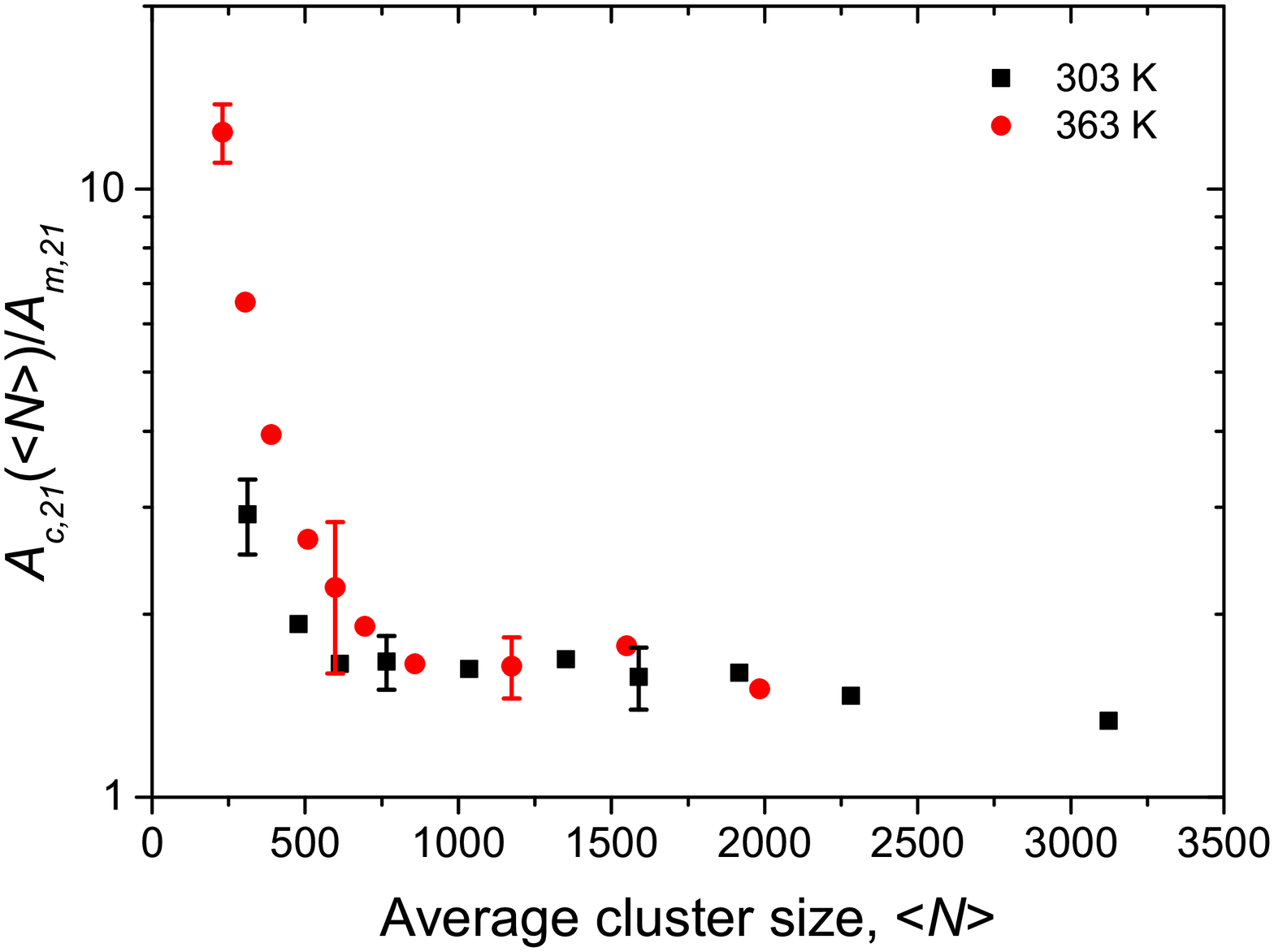}
\caption{Enhancement of the single-atom response in clusters vs. monomers by the ratio $\frac{A_{c,21}(\av{N})}{A_{m,21}}$ obtained from experiments using Eq.~\eqref{eq:ratio} as a function of the average cluster size, $\av{N}$, for $200\lesssim\av{N}\lesssim3000$ for the two temperatures of 303~K (black squares) and 363~K (red circles).}
\label{fig: ratio_of_single_atom_response}
\end{figure}

As Fig.~\ref{fig: HH21intensity_vs_na_temp} shows an enhanced yield for the mixture of clusters and monomers when the average cluster size is in the range from $\sim200$ to $\sim3000$, we show in Fig.~\ref{fig: ratio_of_single_atom_response} the ratio as calculated using Eq.~\eqref{eq:ratio} for this range of average cluster sizes. Figure~\ref{fig: ratio_of_single_atom_response} shows an enhanced single-atom response for atoms within a cluster that increases when the average cluster size becomes smaller. Note that in the limit of an average cluster size of one, i.e., in case of monomers only, the enhancement should disappear and the ratio, $\frac{A_{c,21}(\av{N})}{A_{m,21}} $, shown in Fig. \ref{fig: ratio_of_single_atom_response} drops back to unity. This figure also shows that the single-atom response for atoms inside a cluster is equal within experimental uncertainty for both temperatures when $\av{N}\gtrsim500$. For $\av{N}\lesssim500$, the single-atom response of atoms inside a cluster seems to be larger for the higher temperature (363~K). However, due to the small liquid mass fraction ($\lesssim 0.1\%$), uncertainties in $\av{N}$ and $g$ itself may be responsible for the difference in $A_{c,21}$ at the two temperatures. Moreover, the two temperatures may lead to different cluster size  distributions with the same average cluster size with a corresponding difference in distribution of single-atom responses of the atoms within clusters. The average single-atom response for atoms within a cluster is a factor of 2 to 10 larger than the single-atom response for monomers when $\av{N}\lesssim 500$. An enhanced single-atom response was also observed by Park et al.~\cite{Park2014}, who reported a growth of a single order harmonic yield with $n_a^5$ when using a cluster source with $\av{N}\lesssim 700$. The growth rate we observe in our experiment is less (cf. Fig.~\ref{fig: HH21intensity_vs_na_temp}), which we expect is due to the different liquid mass fraction in the jets produced by the different geometry of the nozzles. As can be seen in Fig.~\ref{fig: ratio_of_single_atom_response}, the enhancement of the single-atom response strongly depends on the average cluster size for small clusters, and a high liquid mass fraction is needed in this range to make full use of this enhancement. In our nozzle, such small clusters are produced only with a small liquid mass fraction at low total atomic number densities and therefore the observed enhancement is rather small.
\section{Conclusion}
We have investigated high-order harmonic generation in a supersonic argon gas jet. To identify the contributions of the generated high-order harmonics from \emph{both} clusters and gas monomers, we measured the harmonic spectra over a broad range of the total atomic number densities (from $3\times10^{16}\;\rm{cm^{-3}}$ to $3\times10^{18}\;\rm{cm^{-3}}$) in the jet at two different reservoir temperatures (303 K and 363 K). 
For the first time in the evaluation of the harmonic yield produced by a mixture of clusters and monomers in such measurements, the detailed variation of the liquid mass fraction, $g$, with pressure and temperature is taken into consideration. We determine this fraction and find, consistently, low values of $g$ below 20\%, within our range of experimental parameters. Changing the temperature of the nozzle, we studied the dependence of the HH yield on $g$, which corresponds to a particular cluster distribution with average size, $\av{N}$, for various total atomic number densities. We use a simple model, which includes macroscopic effects due to absorption and phase matching, to calculate the yield of the $21^{st}$ harmonic order generated by pure argon monomers. Comparing this with the experimental yield for the same harmonic order from the cluster jet and at the same total atomic number density $n_a$, we find that the single-atom response for atoms inside a cluster is enhanced by a factor of up to 10 when $\av{N}\lesssim3000$, while no enhancement is found for larger average cluster sizes. 
We also observe no change of the cut-off energy in the measured harmonic spectra, which indicates that the single-atom three-step model is still applicable for HHG in clusters, i.e., the tunnel ionized electrons collide with their parent ion.
We conclude that using a supersonic gas jet to provide clusters as the nonlinear medium, does  promise a higher harmonic yield via an increased nonlinearity as compared to a gas jet of monomers when the average cluster size is less than 500. To fully exploit this enhancement for high-order harmonic generation, the nozzle should be designed to create smaller clusters (low $\av{N}$) simultaneously with a high liquid-mass fraction ($g\sim1$). Furthermore, the use of cluster jets in high-order harmonic generation could also introduce a density modulation for pursuing a higher yield via quasi-phase matching. Such density modulation can be obtained, e.g., by placing an array of wire-obstacles on the top of our slit nozzle~\cite{Tao2017}.
\ack
This research was supported by the Dutch Technology Foundation STW, which is part of the
Netherlands Organization for Scientific Research (NWO), and partly funded by the Ministry
of Economic Affairs (Project No. 10759).

\section*{Reference}
\bibliographystyle{iopart-num}
\bibliography{HHG_in_clusters}

\end{document}